\documentclass[twocolumn]{aastex6}
\usepackage{amsmath}

\begin{document}

\title{Exclusion of Stellar Companions to Exoplanet Host Stars}

\author{
  Justin M. Wittrock\altaffilmark{1},
  Stephen R. Kane\altaffilmark{1,2},
  Elliott P. Horch\altaffilmark{3},
  Steve B. Howell\altaffilmark{4},
  David R. Ciardi\altaffilmark{5},
  Mark E. Everett\altaffilmark{6}
}
\altaffiltext{1}{Department of Physics \& Astronomy, San Francisco
  State University, 1600 Holloway Avenue, San Francisco, CA 94132,
  USA}
\altaffiltext{2}{Department of Earth Sciences, University of
  California, Riverside, CA 92521, USA}
\altaffiltext{3}{Department of Physics, Southern Connecticut State
  University, New Haven, CT 06515, USA}
\altaffiltext{4}{NASA Ames Research Center, Moffett Field, CA 94035,
  USA}
\altaffiltext{5}{NASA Exoplanet Science Institute, Caltech, MS 100-22,
  770 South Wilson Avenue, Pasadena, CA 91125, USA}
\altaffiltext{6}{National Optical Astronomy Observatory, 950 N. Cherry
  Ave, Tucson, AZ 85719, USA}


\begin{abstract}

Given the frequency of stellar multiplicity in the solar neighborhood,
it is important to study the impacts this can have on exoplanet
properties and orbital dynamics. There have been numerous imaging
survey projects established to detect possible low-mass stellar
companions to exoplanet host stars. Here we provide the results from a
systematic speckle imaging survey of known exoplanet host stars. In
total, 71 stars were observed at 692~nm and 880~nm bands using the
Differential Speckle Survey Instrument (DSSI) at the Gemini-North
Observatory. Our results show that all but 2 of the stars included in
this sample have no evidence of stellar companions with luminosities
down to the detection and projected separation limits of our
instrumentation. The mass-luminosity relationship is used to estimate
the maximum mass a stellar companion can have without being
detected. These results are used to discuss the potential for further
radial velocity follow-up and interpretation of companion signals.

\end{abstract}

\keywords{planetary systems -- techniques: high angular resolution}


\section{Introduction}

The discovery of exoplanets several decades ago (e.g.,
\citet{lat89,lat12,wol92,may95}) has led to a burgeoning and diverse
field of study. A major effort of this work is directed at
characterizing the individual exoplanets and their host stars. For
example, determining the binarity of the host stars has become a
crucial step in understanding exoplanetary systems since the presence
of a binarity companion can have a profound effect on detection
methods and formation scenarios. This is particularly important since
roughly half of all sun-like stars in the solar neighborhood are part
of a multiple-star systems \citep{raghavan10,horch14}. Indeed, the
pursuit of {\it Kepler} candidates \citep{eve15,kra16} and Robo-AO
observations of radial velocity (RV) exoplanet host stars has
significantly contributed to our knowledge of this large rate of
stellar multiplicity.

The presence of binary companions can considerably affect stellar
measurements intended to discover and/or characterize exoplanets and
cause severe blended contamination for transit observations
\citep{car15,cia15,gil15}. One of the main consequences of this for
exoplanets detected by the transit method is the underestimation of
their planetary radii determined from the depth of the planetary
transit \citep{cia15,furlan17}. For systems discovered using the RV
method, the presence of stellar companions can manifest as linear
trends in the data, the precise origins of which can remain unresolved
due to the insufficient time baseline and the $\sin i$ ambiguity of
the companion mass interpretation \citep{cre12,kan14}. The range of
possible planetary formation scenarios is inhibited by the
multiplicity of the stars, and can impact such aspects as the orbital
stability of the planets \citep{holman99}. Furthermore, there has been
a noted effect of stellar binarity on masses, orbital periods
\citep{zucker02}, and eccentricities \citep{eggenberger04} of planets
in such binary systems. Consequently, it has become critical for us to
establish the multiplicity of exoplanet host stars so that we can be
absolutely confident with the resulting interpretation of exoplanet
signals and be successful in fully characterizing the overall system
properties.

\citet{wittrock16} described a survey for exoplanet host star
multiplicity and presented the detection of stellar companions to 2 of
71 surveyed exoplanet host stars, HD~2638 and HD~164509. Here we
present the results for the remaining 69 stars of the survey that
place significant constraints on the presence of stellar companions to
those stars. In Section \ref{obs}, we discuss the method of detection,
the range of targets that were selected for analysis, and the
properties of null-detection systems. Section \ref{red} briefly
reviews the details of the data reduction and includes sensitivity
plots of the observed systems. Section \ref{res} presents the results
from the data analysis, and Section \ref{con} provides discussion of
further work and concluding remarks.


\section{Selections and Properties of the Targeted Systems}
\label{obs}

A large survey project was established to search for stellar
companions to a subset of the known RV exoplanet host stars. In total
71 stars were observed in July 2014 using the Differential Speckle
Survey Instrument, or DSSI \citep{horch09}; that instrument was
stationed at the Gemini-North Observatory at the time of
observations. The stars were selected from the known RV exoplanet host
star population where there was no known stellar companion. Two of the
stars, HD~2638 and HD~164509, were found to have evidence for bound
stellar companions contained in the data \citep{wittrock16}. Tables
\ref{propi} and \ref{propii} list the 69 targets from the survey for
which no stellar companion was detected. The first table includes
spectral types, apparent magnitudes m$_V$, proper motions (denoted as
$\mu$), parallaxes, distances, and the number of exoplanets each star
hosts, while the second table tallies stellar masses, radii,
luminosity, effective temperatures, surface gravity, age, and
metallicity. The data within both tables were taken from multiple
literature and exoplanet databases (see reference section). The DSSI
used two different filters, 692~nm and 880~nm, to acquire the speckle
images of those targets. The 692~nm filter has FWHM of 40~nm, and the
880~nm filter has FWHM of 50~nm. All images were reduced using a data
reduction pipeline, the details of which are provided in
Section~\ref{red}. Afterward, the images were examined by eye and also
using the speckle reduced data plots for any companion source
appearing next to the target.

\begin{figure*}
  \begin{center}
    \begin{tabular}{cc}
      \includegraphics[width=8cm]{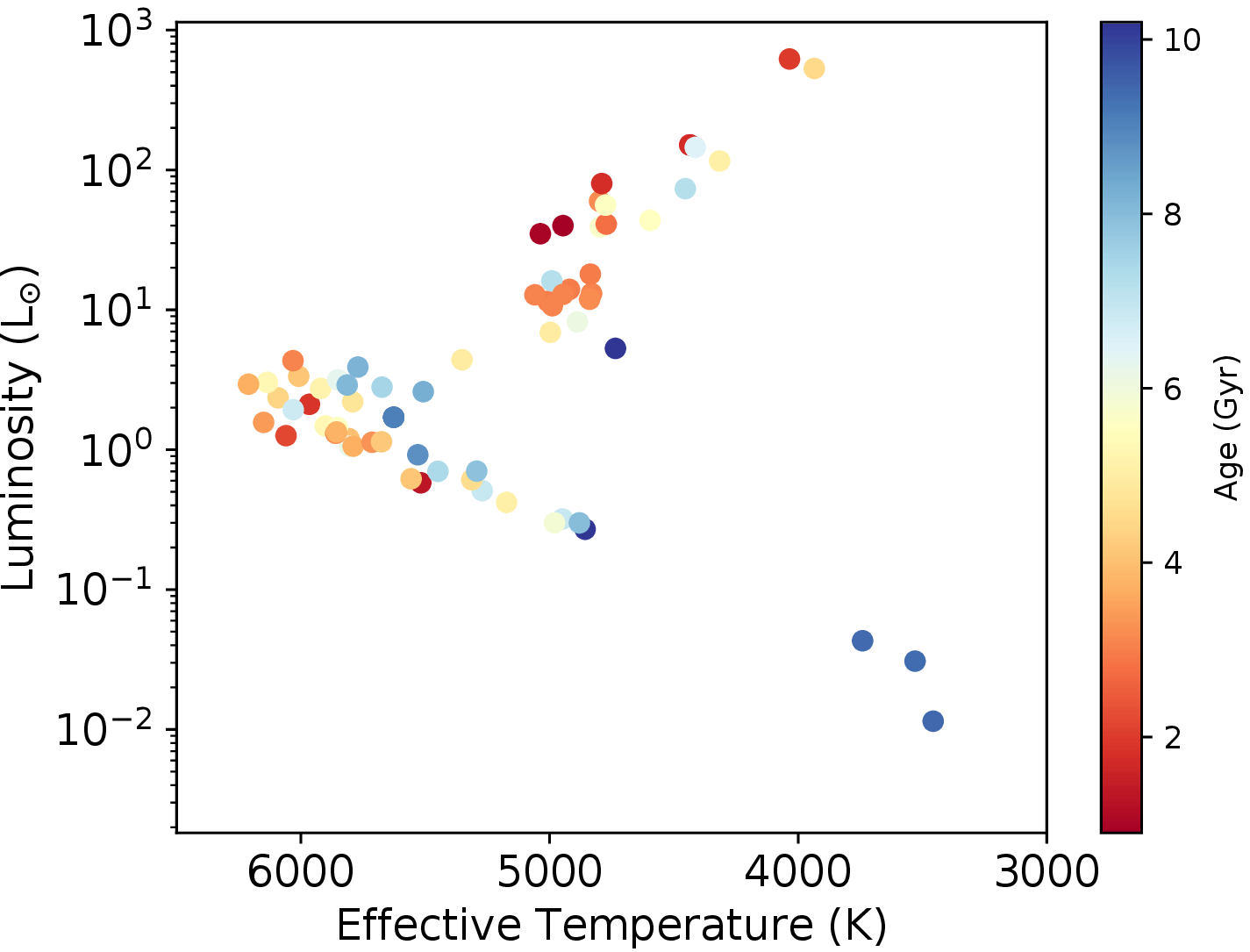} &
      \includegraphics[width=8cm]{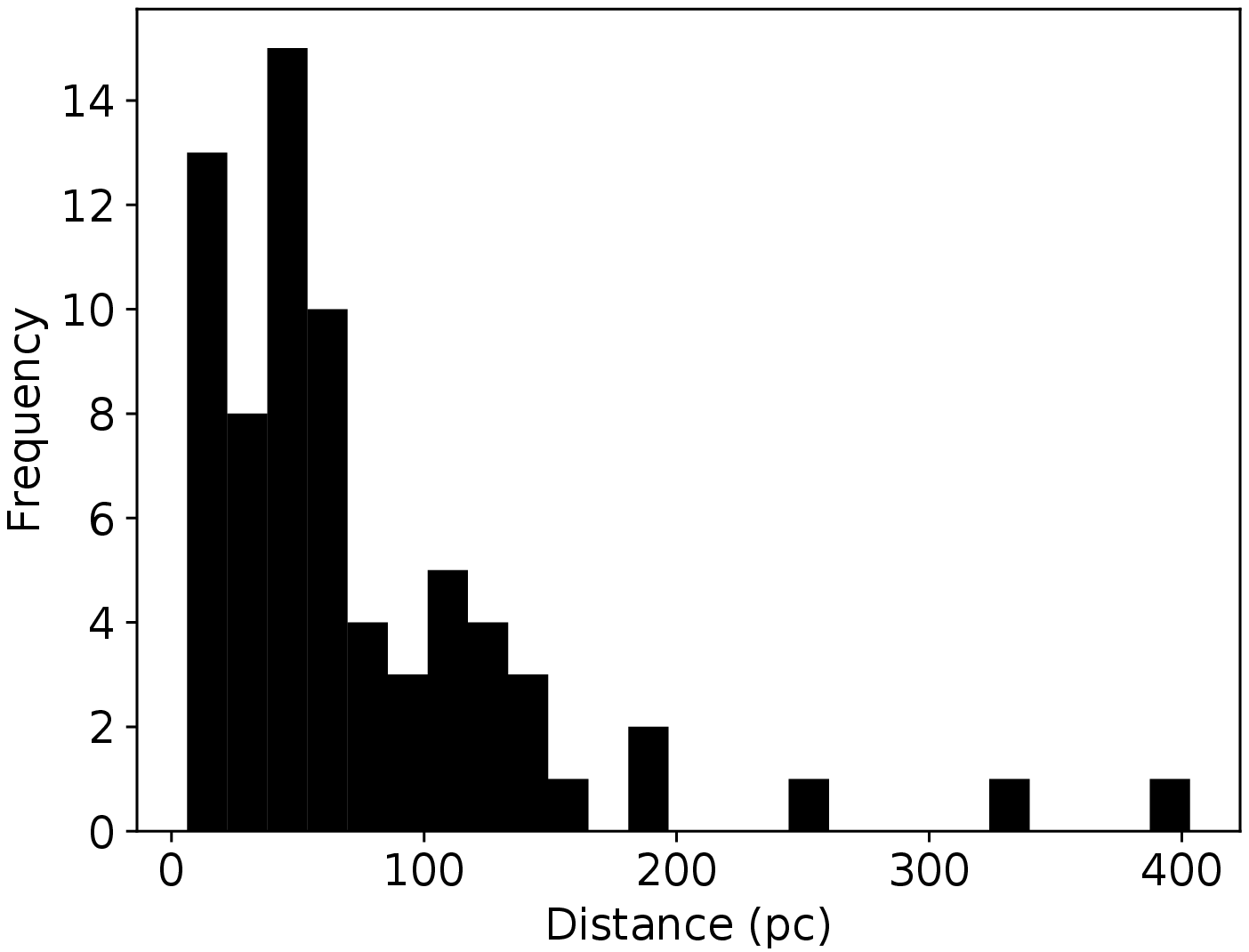}
    \end{tabular}
  \end{center}
  \caption{Left: H-R diagram of the stars included in the survey,
    where the color bar indicates the stellar age in Gyr. Right:
    Histogram of distances to the surveyed stars. The data from these
    plots are from Tables \ref{propi} and \ref{propii}.}
  \label{tableplot}
\end{figure*}

Figure~\ref{tableplot} showcases the Hertzsprung-Russell (H-R) diagram
and distance histogram of the survey using the data from Tables
\ref{propi} and \ref{propii}. BD+48~738 and HD~13189 were excluded
since their luminosity and/or age data are unavailable, and HD~240237
was excluded due to its extremely large distance of 5300~pc. As
described by \citet{wittrock16}, this is a magnitude-limited survey
that targets the brightest of known exoplanet host stars, and so the
sample consists mostly of relatively nearby dwarf stars with a peak in
the distance distribution of $\sim$50~pc. The large distances of the
giant stars results in a small angular separation sensitivity for
detecting stellar companions.

\begin{deluxetable*}{lcCCCCCc}
  \tabletypesize{\scriptsize}
  \tablecolumns{8}
  \tablewidth{0pt}
  \tablecaption{\label{propi} Stellar Properties of RV Exoplanet Host Stars with Null-Detections I}
  \tablehead{
    Star &
    \colhead{Spectral Type} &
    \colhead{m$_V$\tablenotemark{(33)}} &
    \colhead{$\mu$ $\alpha$, $\delta$ ($mas/yr$)\tablenotemark{(33)}} &
    \colhead{Parallax (mas)\tablenotemark{(33)}} &
    \colhead{Distance (pc)} &
    \colhead{Planets} &
    \colhead{References}
    }
  \startdata
BD+14 4559	&	K2 V	&	9.7768	&	235.80, 1.78		&	20.68 \pm 1.24	&	48.36 \pm 2.90			&	1	&	(25)			\\
BD+48 738	&	K0 III	&	9.14	&	3.7, -6.5			&	2.85 \pm 0.00	&	350.88 \pm 0.00			&	1	&	(10),(28),(36)	\\
GJ 581		&	M3 V	&	10.5759	&	-1227.67, -97.78	&	160.91 \pm 2.62	&	6.21 \pm 0.10			&	3	&	(2)				\\
GJ 649		&	M2 V	&	9.7165	&	-114.07, -506.26	&	96.67 \pm 1.39	&	10.34 \pm 0.15			&	2	&	(35)			\\
GJ 849		&	M3 V	&	10.3672	&	1130.27, -19.27		&	109.94 \pm 2.07	&	9.10 \pm 0.17			&	1	&	(3)				\\
HD 1461		&	G3 V	&	6.6029	&	416.87, -143.83		&	43.02 \pm 0.51	&	23.25 \pm 0.28			&	2	&	(14)			\\
HD 1502		&	K0 IV	&	8.5196	&	74.64, -18.15		&	6.28 \pm 0.75	&	159.24 \pm 19.02		&	1	&	(6)				\\
HD 3651		&	K0 V	&	6.03	&	-461.32, -370.02	&	90.42 \pm 0.32	&	11.06 \pm 0.04			&	2	&	(14)			\\
HD 4313		&	G5 IV	&	7.9939	&	-5.14, 6.69			&	7.3 \pm 0.76	&	136.99 \pm 14.26		&	1	&	(6)				\\
HD 5319		&	K3 IV	&	8.2069	&	-4.93, -49.66		&	8.74 \pm 0.86	&	114.42 \pm 11.26		&	2	&	(12)			\\
HD 5891		&	G5 III	&	8.2541	&	2.64, -41.89		&	3.98 \pm 1.21	&	251.26 \pm 76.39		&	1	&	(6)				\\
HD 6718		&	G5 V	&	8.5834	&	192.24, 19.77		&	18.23 \pm 0.76	&	54.85 \pm 2.29			&	1	&	(23)			\\
HD 7449		&	F8 V	&	7.6205	&	-160.79, -138.95	&	25.69 \pm 0.48	&	38.93 \pm 0.73			&	2	&	(4)				\\
HD 8574		&	F8 V	&	7.2497	&	250.87, -158.06		&	22.44 \pm 0.53	&	44.56 \pm 1.05			&	1	&	(6)				\\
HD 9446		&	G5 V	&	8.5125	&	192.01, -53.99		&	19.1 \pm 1.06	&	52.36 \pm 2.91			&	2	&	(6)				\\
HD 10697	&	G5 IV	&	6.4169	&	-44.75, -105.35		&	30.7 \pm 0.43	&	32.57 \pm 0.46			&	1	&	(6)				\\
HD 12661	&	G6 V	&	7.567	&	-107.12, -174.69	&	28.61 \pm 0.61	&	34.95 \pm 0.75			&	2	&	(8)				\\
HD 13189	&	K2 II	&	7.6968	&	2.62, 5.32			&	1.78 \pm 0.73	&	561.80 \pm 230.40		&	1	&	(17)			\\
HD 13931	&	G0 V	&	7.7426	&	99.03, -183.19		&	22.61 \pm 0.66	&	44.23 \pm 1.29			&	1	&	(16)			\\
HD 16175	&	G0 V	&	7.4156	&	-38.90, -40.37		&	17.28 \pm 0.67	&	57.87 \pm 2.24			&	1	&	(6)				\\
HD 16400	&	G5 III	&	5.8154	&	40.18, -42.91		&	10.81 \pm 0.45	&	92.51 \pm 3.85			&	1	&	(6)				\\
HD 16760	&	G5 V	&	8.8411	&	79.20, -107.49		&	22 \pm 2.35		&	45.45 \pm 4.86			&	1	&	(30)			\\
HD 17092	&	K0 III	&	7.73	&	37.9, -13.6			&	9.2 \pm 5.5		&	108.70 \pm 64.98		&	1	&	(6),(18),(24)	\\
HD 136118	&	F9 V	&	7.0513	&	-122.69, 23.72		&	21.47 \pm 0.54	&	46.58 \pm 1.17			&	1	&	(7)				\\
HD 136418	&	G5 IV	&	8.0279	&	-19.66, -181.92		&	10.18 \pm 0.58	&	98.23 \pm 5.60			&	1	&	(6)				\\
HD 137510	&	G0 IV	&	6.3856	&	-54.91, -5.39		&	24.24 \pm 0.51	&	41.25 \pm 0.87			&	1	&	(5)				\\
HD 139357	&	K4 III	&	6.1335	&	-18.32, 1.64		&	8.47 \pm 0.3	&	118.06 \pm 4.18			&	1	&	(6)				\\
HD 142245	&	K0 IV	&	7.6302	&	-55.58, -20.82		&	9.13 \pm 0.62	&	109.53 \pm 7.44			&	1	&	(6)				\\
HD 143107	&	K3 III	&	4.2992	&	-77.07, -60.61		&	14.73 \pm 0.21	&	67.89 \pm 0.97			&	1	&	(6)				\\
HD 143761	&	G0 V	&	5.5246	&	-196.63, -773.02	&	58.02 \pm 0.28	&	17.24 \pm 0.08			&	2	&	(32)			\\
HD 145457	&	K0 III	&	6.7416	&	-18.34, 36.89		&	7.98 \pm 0.45	&	125.31 \pm 7.07			&	1	&	(6)				\\
HD 145675	&	K0 V	&	6.7595	&	131.83, -297.54		&	56.91 \pm 0.34	&	17.57 \pm 0.10			&	1	&	(14)			\\
HD 149143	&	G0 IV	&	8.0354	&	-9.26, -87.31		&	16.12 \pm 0.83	&	62.03 \pm 3.19			&	1	&	(9)				\\
HD 152581	&	K0 IV	&	8.5372	&	11.49, -15.79		&	5.39 \pm 0.96	&	185.53 \pm 33.04		&	1	&	(6)				\\
HD 154345	&	G8 V	&	6.907	&	123.27, 853.63		&	53.8 \pm 0.32	&	18.59 \pm 0.11			&	1	&	(6)				\\
HD 155358	&	G0 V	&	7.3946	&	-222.45, -215.97	&	22.67 \pm 0.48	&	44.11 \pm 0.93			&	2	&	(6)				\\
HD 156279	&	K0 V	&	8.2107	&	-1.21, 161.21		&	27.32 \pm 0.44	&	36.60 \pm 0.59			&	1	&	(6)				\\
HD 156668	&	K3 V	&	8.5711	&	-71.16, 217.36		&	40.86 \pm 0.86	&	24.47 \pm 0.52			&	1	&	(14)			\\
HD 158038	&	K2 II	&	7.6439	&	48.35, -59.04		&	9.65 \pm 0.74	&	103.63 \pm 7.95			&	1	&	(6)				\\
HD 163607	&	G5 IV	&	8.1487	&	-75.74, 120.05		&	14.53 \pm 0.46	&	68.82 \pm 2.18			&	2	&	(11)			\\
HD 164922	&	G9 V	&	7.151	&	389.41, -602.03		&	45.21 \pm 0.54	&	22.12 \pm 0.26			&	2	&	(14)			\\
HD 167042	&	K1 IV	&	6.1356	&	107.94, 247.35		&	19.91 \pm 0.26	&	50.23 \pm 0.66			&	1	&	(20)			\\
HD 170693	&	K2 III	&	4.9835	&	105.83, -27.24		&	10.36 \pm 0.2	&	96.53 \pm 1.86			&	1	&	(6)				\\
HD 171028	&	G0 IV	&	8.31	&	-43.8, -13.4		&	9.1 \pm 7.8		&	109.89 \pm 94.19		&	1	&	(6),(18),(27)	\\
HD 173416	&	G8 III	&	6.2114	&	21.11, 58.23		&	7.17 \pm 0.28	&	139.47 \pm 5.45			&	1	&	(6)				\\
HD 177830	&	K0 IV	&	7.3455	&	-40.84, -51.75		&	16.94 \pm 0.63	&	59.03 \pm 2.20			&	2	&	(34)			\\
HD 180314	&	K0 III	&	6.7743	&	47.19, 19.71		&	7.61 \pm 0.39	&	131.41 \pm 6.73			&	1	&	(31)			\\
HD 187123	&	G2 V	&	7.9689	&	143.18, -123.91		&	20.72 \pm 0.53	&	48.26 \pm 1.23			&	2	&	(13)			\\
HD 190228	&	G5 IV	&	7.452	&	105.2, -69.82		&	16.23 \pm 0.64	&	61.61 \pm 2.43			&	1	&	(6)				\\
HD 192263	&	K2.5 V	&	7.931	&	-61.13, 261.37		&	51.77 \pm 0.78	&	19.32 \pm 0.29			&	1	&	(14)			\\
HD 197037	&	F7 V	&	6.9226	&	-62.47, -220.96		&	30.93 \pm 0.38	&	32.33 \pm 0.40			&	1	&	(26)			\\
HD 199665	&	G6 III	&	5.6682	&	-48.75, -34.43		&	13.28 \pm 0.31	&	75.30 \pm 1.76			&	1	&	(6)				\\
HD 200964	&	K0 IV	&	6.6386	&	94.99, 50.47		&	13.85 \pm 0.52	&	72.20 \pm 2.71			&	2	&	(29)			\\
HD 206610	&	K0 IV	&	8.5066	&	2.35, 2.34			&	5.16 \pm 0.95	&	193.80 \pm 35.68		&	1	&	(6)				\\
HD 208527	&	M1 III	&	6.4842	&	2.00, 15.30			&	2.48 \pm 0.38	&	403.23 \pm 61.78		&	1	&	(22)			\\
HD 210277	&	G8 V	&	6.6823	&	85.07, -449.74		&	46.38 \pm 0.48	&	21.56 \pm 0.22			&	1	&	(14)			\\
HD 210702	&	K1 IV	&	6.0932	&	-3.15, -18.02		&	18.2 \pm 0.39	&	54.95 \pm 1.18			&	1	&	(19)			\\
HD 217014	&	G2 V	&	5.5865	&	207.25, 60.34		&	64.07 \pm 0.38	&	15.61 \pm 0.09			&	1	&	(15)			\\
HD 217107	&	G8 IV	&	6.3124	&	-6.35, -15.80		&	50.36 \pm 0.38	&	19.86 \pm 0.15			&	2	&	(14)			\\
HD 217786	&	F9 V	&	7.9103	&	-88.78, -170.13		&	18.23 \pm 0.72	&	54.85 \pm 2.17			&	1	&	(6)				\\
HD 218566	&	K3 V	&	8.7269	&	632.56, -97.02		&	35.02 \pm 1.14	&	28.56 \pm 0.93			&	1	&	(6)				\\
HD 219828	&	G0 IV	&	8.1795	&	-4.15, 4.14			&	13.83 \pm 0.74	&	72.31 \pm 3.87			&	2	&	(6)				\\
HD 220074	&	M2 III	&	6.4885	&	7.68, -5.43			&	3.08 \pm 0.43	&	324.68 \pm 45.33		&	1	&	(21)			\\
HD 220773	&	F9 V	&	7.2306	&	26.90, -222.87		&	19.65 \pm 0.65	&	50.89 \pm 1.68			&	1	&	(26)			\\
HD 221345	&	K0 III	&	5.3841	&	286.72, -84.22		&	12.63 \pm 0.27	&	79.18 \pm 1.69			&	1	&	(6)				\\
HD 222155	&	G2 V	&	7.2445	&	195.33, -117.13		&	20.38 \pm 0.62	&	49.07 \pm 1.49			&	1	&	(1)				\\
HD 231701	&	F8 V	&	9.0929	&	63.85, 16.46		&	8.44 \pm 1.05	&	118.48 \pm 14.74		&	1	&	(6)				\\
HD 240210	&	K3 III	&	8.33	&	18.0, 7.9			&	7 \pm 2.6		&	142.86 \pm 53.06		&	1	&	(18),(25)		\\
HD 240237	&	K2 III	&	8.2959	&	-0.74, -5.13		&	0.19 \pm 0.72	&	5263.16 \pm 19944.60	&	1	&	(10)
  \enddata
~~(1) {\citet{boisse12}}, (2) {\citet{bonfils05}}, (3) {\citet{bonfils13}}, (4) {\citet{dumusque11}}, (5) {\citet{endl04}}, (6) {\citet{esa97}}, (7) {\citet{fischer02}}, (8) {\citet{fischer03}}, (9) {\citet{fischer06}}, (10) {\citet{gettel12}}, (11) {\citet{giguere12}}, (12) {\citet{giguere15}}, (13) {\citet{gray01}}, (14) {\citet{gray03}}, (15) {\citet{gray06}}, (16) {\citet{grenier99}}, (17) {\citet{hatzes05}}, (18) {\citet{hog00}}, (19) {\citet{johnson07}}, (20) {\citet{johnson08}}, (21) {\citet{kidger03}}, (22) {\citet{lee13}}, (23) {\citet{naef10}}, (24) {\citet{niedzielski07}}, (25) {\citet{niedzielski09}}, (26) {\citet{robertson12}}, (27) {\citet{santos07}}, (28) {\citet{santos13}}, (29) {\citet{santos15}}, (30) {\citet{sato09}}, (31) {\citet{sato10}}, (32) {\citet{belle09}}, (33) {\citet{leeuwen07}}, (34) {\citet{vogt00}}, (35) {\citet{braun14}}, (36) {\citet{zacharias04}}
\end{deluxetable*}

\begin{deluxetable*}{lCCCCCCCc}
  \tabletypesize{\scriptsize}
  \tablecolumns{9}
  \tablewidth{0pt}
  \tablecaption{\label{propii} Stellar Properties of RV Exoplanet Host Stars with Null-Detection II}
  \tablehead{
    Name &
    \colhead{$M_{\star}$ ($M_{\odot}$)} &
    \colhead{$R_{\star}$ ($R_{\odot}$)} &
    \colhead{$L_{\star}$ ($L_{\odot}$)} &
    \colhead{$T_e$ (K)} &
    \colhead{$\log$ $g$ ($cm/s^2$)} &
    \colhead{Age (Gyr)} &
    \colhead{$[$Fe/H$]$} &
    \colhead{References}
    }
  \startdata
BD+14 4559	&	0.82 \pm 0.02	&	0.78 \pm 0.02		&	0.32 \pm 0.01		&	4948 \pm 25		&	4.57 \pm 0.03	&	6.9 \pm 4.2		&	0.17 \pm 0.06	&	(3),(15)	\\
BD+48 738	&	0.74 \pm 0.39	&	11 \pm 1			&	49 \pm 37.2			&	4519 \pm 30		&	2.51 \pm 0.03	&	 -				&	-0.24 \pm 0.02	&	(5),(8)	\\
GJ 581		&	0.306 \pm 0.011	&	0.299 \pm 0.007		&	0.01146 \pm 0.00061	&	3457 \pm 22		&	4.96 \pm 0.25	&	9.44 \pm 0.58	&	-0.15 \pm 0.08	&	(12),(15)	\\
GJ 649		&	0.527 \pm 0.013	&	0.495 \pm 0.012		&	0.04308 \pm 0.00276	&	3741 \pm 39		&	4.76 \pm 0.12	&	9.42 \pm 0.57	&	0.03 \pm 0.08	&	(12),(15)	\\
GJ 849		&	0.482 \pm 0.048	&	0.47 \pm 0.018		&	0.03079 \pm 0.00315	&	3530 \pm 60		&	4.8 \pm 0.14	&	9.4 \pm 0.58	&	0.37 \pm 0.08	&	(12),(15)	\\
HD 1461		&	1.07 \pm 0.01	&	1.08 \pm 0.01		&	1.2 \pm 0.01		&	5807 \pm 20		&	4.39 \pm 0.01	&	4 \pm 0.7		&	0.16 \pm 0.03	&	(3),(11)	\\
HD 1502		&	1.46 \pm 0.04	&	4.5 \pm 0.1			&	11.5 \pm 0.2		&	5006 \pm 25		&	3.29 \pm 0.02	&	3 \pm 0.3		&	-0.01 \pm 0.06	&	(3),(8)	\\
HD 3651		&	0.88 \pm 0.02	&	0.86 \pm 0.01		&	0.51 \pm 0.01		&	5271 \pm 26		&	4.51 \pm 0.02	&	6.9 \pm 2.8		&	0.19 \pm 0.02	&	(3),(11)	\\
HD 4313		&	1.49 \pm 0.04	&	5.2 \pm 0.1			&	14 \pm 0.2			&	4920 \pm 21		&	3.18 \pm 0.02	&	3 \pm 0.3		&	0.11 \pm 0.07	&	(3),(8)	\\
HD 5319		&	1.2 \pm 0.1		&	4 \pm 0.1			&	8.2 \pm 0.1			&	4888 \pm 39		&	3.3 \pm 0.04	&	6.1 \pm 1.4		&	0.15 \pm 0.03	&	(3),(6)	\\
HD 5891		&	1.1 \pm 0.1		&	9.1 \pm 0.2			&	39.1 \pm 0.4		&	4796 \pm 41		&	2.57 \pm 0.05	&	5.7 \pm 1.5		&	-0.37 \pm 0.04	&	(3),(8)	\\
HD 6718		&	0.97 \pm 0.02	&	1.02 \pm 0.03		&	1.06 \pm 0.02		&	5805 \pm 46		&	4.4 \pm 0.03	&	6.2 \pm 2		&	-0.11 \pm 0.05	&	(1),(3)	\\
HD 7449		&	1.05 \pm 0.02	&	1.02 \pm 0.02		&	1.26 \pm 0.02		&	6060 \pm 42		&	4.44 \pm 0.02	&	2.2 \pm 1.3		&	-0.11 \pm 0.01	&	(3),(15)	\\
HD 8574		&	1.17 \pm 0.02	&	1.38 \pm 0.04		&	2.35 \pm 0.04		&	6092 \pm 56		&	4.22 \pm 0.03	&	4.4 \pm 0.6		&	0.06 \pm 0.07	&	(3),(15)	\\
HD 9446		&	1.04 \pm 0.03	&	1.03 \pm 0.03		&	1.06 \pm 0.03		&	5790 \pm 45		&	4.43 \pm 0.03	&	3.7 \pm 2		&	0.09 \pm 0.05	&	(3),(15)	\\
HD 10697	&	1.12 \pm 0.01	&	1.7 \pm 0.1			&	2.8 \pm 0.04		&	5674 \pm 93		&	4 \pm 0.03		&	7.5 \pm 0.4		&	0.15 \pm 0.04	&	(3),(8)	\\
HD 12661	&	1.09 \pm 0.01	&	1.08 \pm 0.01		&	1.13 \pm 0.01		&	5714 \pm 22		&	4.4 \pm 0.01	&	3.3 \pm 0.6		&	0.36 \pm 0.05	&	(3),(15)	\\
HD 13189	&	1.08 \pm 0.17	&	 -	 				&	 -	 				&	4228 \pm 242	&	2.09 \pm 0.61	&	 -	 			&	-0.5 \pm 0.14	&	(15)	\\
HD 13931	&	1.07 \pm 0.02	&	1.17 \pm 0.03		&	1.48 \pm 0.03		&	5902 \pm 52		&	4.33 \pm 0.03	&	5.3 \pm 1.3		&	0.07 \pm 0.01	&	(3),(17)	\\
HD 16175	&	1.3 \pm 0.05	&	1.69 \pm 0.03		&	3.35 \pm 0.02		&	6009 \pm 44		&	4.09 \pm 0.02	&	4.1 \pm 0.8		&	0.37 \pm 0.03	&	(3),(16)	\\
HD 16400	&	1.4 \pm 0.1		&	11.2 \pm 0.2		&	59.8 \pm 0.4		&	4799 \pm 24		&	2.49 \pm 0.03	&	3.2 \pm 0.5		&	0 \pm 0.04		&	(3),(8)	\\
HD 16760	&	0.93 \pm 0.01	&	0.835 \pm 0.005		&	0.58 \pm 0.002		&	5518 \pm 11		&	4.56 \pm 0.01	&	1.3 \pm 0.9		&	0 \pm 0.02		&	(2),(14)	\\
HD 17092	&	1.246 \pm 0.179	&	10.439 \pm 1.31		&	43.64 \pm 11.23		&	4596 \pm 65		&	2.45 \pm 0.17	&	5.58 \pm 2.669	&	0.05 \pm 0.04	&	(16)	\\
HD 136118	&	1.15 \pm 0.03	&	1.54 \pm 0.03		&	3.03 \pm 0.01		&	6135 \pm 37		&	4.12 \pm 0.03	&	5.3 \pm 0.6		&	-0.01 \pm 0.053	&	(2),(7)	\\
HD 136418	&	1.2 \pm 0.1		&	3.5 \pm 0.1			&	6.9 \pm 0.1			&	4997 \pm 40		&	3.43 \pm 0.04	&	5 \pm 1			&	-0.09 \pm 0.03	&	(3),(16)	\\
HD 137510	&	1.41 \pm 0.01	&	1.91 \pm 0.03		&	4.33 \pm 0.01		&	6032 \pm 44		&	4.02 \pm 0.02	&	3.1 \pm 0.2		&	0.29 \pm 0.12	&	(2),(9)	\\
HD 139357	&	1.1 \pm 0.1		&	14.4 \pm 0.4		&	73.5 \pm 1.3		&	4454 \pm 39		&	2.2 \pm 0.1		&	7.2 \pm 1.8		&	0.19 \pm 0.05	&	(3),(16)	\\
HD 142245	&	1.52 \pm 0.05	&	5.2 \pm 0.1			&	13.1 \pm 0.2		&	4831 \pm 28		&	3.19 \pm 0.03	&	3.1 \pm 0.3		&	0.23 \pm 0.03	&	(3),(15)	\\
HD 143107	&	1.44 \pm 0.18	&	21 \pm 0			&	151 \pm 0			&	4436 \pm 56		&	1.94 \pm 0.15	&	1.74 \pm 0.37	&	-0.22 \pm 0.03	&	(13),(15)	\\
HD 143761	&	0.889 \pm 0.03	&	1.3617 \pm 0.0262	&	1.706 \pm 0.042		&	5627 \pm 54		&	4.121 \pm 0.018	&	9.1 \pm 1		&	-0.31 \pm 0.05	&	(2),(4)	\\
HD 145457	&	1.5 \pm 0.1		&	9.4 \pm 0.2			&	41 \pm 1			&	4772 \pm 45		&	2.66 \pm 0.05	&	2.8 \pm 0.6		&	-0.13 \pm 0.03	&	(3),(16)	\\
HD 145675	&	0.97 \pm 0.01	&	0.93 \pm 0.01		&	0.61 \pm 0.01		&	5313 \pm 18		&	4.48 \pm 0.02	&	4.6 \pm 1.5		&	0.5 \pm 0.06	&	(3),(11)	\\
HD 149143	&	1.21 \pm 0.03	&	1.5 \pm 0.1			&	2.2 \pm 0.1			&	5792 \pm 58		&	4.17 \pm 0.03	&	4.8 \pm 0.8		&	0.45 \pm 0.07	&	(3),(15)	\\
HD 152581	&	1 \pm 0.1		&	5.4 \pm 0.1			&	16.1 \pm 0.2		&	4991 \pm 45		&	3 \pm 0.1		&	7.2 \pm 2		&	-0.3 \pm 0.02	&	(3),(16)	\\
HD 154345	&	0.9 \pm 0.01	&	0.85 \pm 0.01		&	0.62 \pm 0.002		&	5557 \pm 15		&	4.53 \pm 0.01	&	4.1 \pm 1.2		&	-0.09 \pm 0.02	&	(3),(11)	\\
HD 155358	&	1.1 \pm 0.1		&	1.36 \pm 0.03		&	2.11 \pm 0.02		&	5966 \pm 53		&	4.2 \pm 0.04	&	1.9 \pm 4.5		&	-0.62 \pm 0.02	&	(3),(15)	\\
HD 156279	&	0.93 \pm 0.02	&	0.94 \pm 0.02		&	0.7 \pm 0.01		&	5449 \pm 31		&	4.45 \pm 0.03	&	7.4 \pm 2.2		&	0.14 \pm 0.01	&	(3),(15)	\\
HD 156668	&	0.75 \pm 0.01	&	0.73 \pm 0.01		&	0.27 \pm 0.01		&	4857 \pm 18		&	4.58 \pm 0.01	&	10.2 \pm 2.8	&	-0.04 \pm 0.05	&	(3),(15)	\\
HD 158038	&	1.5 \pm 0.1		&	4.9 \pm 0.1			&	11.9 \pm 0.1		&	4839 \pm 29		&	3.23 \pm 0.03	&	3.2 \pm 0.4		&	0.16 \pm 0.05	&	(3),(16)	\\
HD 163607	&	1.1 \pm 0.02	&	1.8 \pm 0.1			&	2.6 \pm 0.1			&	5508 \pm 15		&	3.98 \pm 0.01	&	8.3 \pm 0.5		&	0.22 \pm 0.02	&	(3),(16)	\\
HD 164922	&	0.874 \pm 0.012	&	0.999 \pm 0.017		&	0.703 \pm 0.017		&	5293 \pm 32		&	4.387 \pm 0.014	&	7.9 \pm 2.7		&	0.16 \pm 0.05	&	(3),(4)	\\
HD 167042	&	1.46 \pm 0.05	&	4.4 \pm 0.1			&	10.7 \pm 0.1		&	4989 \pm 32		&	3.31 \pm 0.03	&	3.1 \pm 0.3		&	-0.01 \pm 0.06	&	(3),(8)	\\
HD 170693	&	1.1 \pm 0.1		&	20.6 \pm 0.6		&	145 \pm 3			&	4414 \pm 40		&	1.8 \pm 0.1		&	6.5 \pm 1.7		&	-0.41 \pm 0.03	&	(3),(16)	\\
HD 171028	&	0.98 \pm 0.04	&	2 \pm 0.2			&	3.9 \pm 0.5			&	5771 \pm 46		&	3.84 \pm 0.03	&	8.2 \pm 1.1		&	-0.47 \pm 0.02	&	(3),(8)	\\
HD 173416	&	1.8 \pm 0.2		&	13 \pm 0.3			&	80 \pm 2			&	4790 \pm 37		&	2.5 \pm 0.1		&	1.8 \pm 0.7		&	-0.15 \pm 0.03	&	(3),(16)	\\
HD 177830	&	1.1 \pm 0.1		&	3.4 \pm 0.1			&	5.3 \pm 0.1			&	4735 \pm 31		&	3.39 \pm 0.04	&	10.2 \pm 1.7	&	0.09 \pm 0.04	&	(3),(8)	\\
HD 180314	&	2.3 \pm 0.1		&	8.7 \pm 0.3			&	40 \pm 1			&	4946 \pm 55		&	2.92 \pm 0.05	&	0.9 \pm 0.2		&	0.11 \pm 0.04	&	(3),(16)	\\
HD 187123	&	1.06 \pm 0.02	&	1.17 \pm 0.03		&	1.44 \pm 0.02		&	5853 \pm 53		&	4.32 \pm 0.03	&	5.6 \pm 1.3		&	0.13 \pm 0.03	&	(3),(15)	\\
HD 190228	&	1.18 \pm 0.04	&	2.4 \pm 0.1			&	4.4 \pm 0.2			&	5352 \pm 30		&	3.73 \pm 0.02	&	5 \pm 0.5		&	-0.24 \pm 0.06	&	(2),(8)	\\
HD 192263	&	0.78 \pm 0.02	&	0.73 \pm 0.01		&	0.3 \pm 0.01		&	4980 \pm 20		&	4.59 \pm 0.02	&	5.9 \pm 3.9		&	-0.01 \pm 0.05	&	(3),(11)	\\
HD 197037	&	1.063 \pm 0.022	&	1.105 \pm 0.023		&	1.568 \pm 0.074		&	6150 \pm 34		&	4.37 \pm 0.04	&	3.408 \pm 0.924	&	-0.16 \pm 0.03	&	(16)	\\
HD 199665	&	2.1 \pm 0.1		&	7.8 \pm 0.3			&	35 \pm 1			&	5037 \pm 57		&	2.98 \pm 0.04	&	1 \pm 0.1		&	0.1 \pm 0.02	&	(3),(8)	\\
HD 200964	&	1.4 \pm 0.1		&	4.7 \pm 0.1			&	12.8 \pm 0.2		&	5059 \pm 34		&	3.23 \pm 0.03	&	3.1 \pm 0.4		&	-0.16 \pm 0.03	&	(3),(8)	\\
HD 206610	&	1.51 \pm 0.05	&	6 \pm 0.2			&	18 \pm 1			&	4836 \pm 30		&	3.05 \pm 0.03	&	3 \pm 0.3		&	0.09 \pm 0.05	&	(3),(8)	\\
HD 208527	&	1.6 \pm 0.4		&	51.1 \pm 8.3		&	621.3 \pm 205.8		&	4035 \pm 65		&	1.4 \pm 0.2		&	2 \pm 1.3		&	-0.09 \pm 0.16	&	(10)	\\
HD 210277	&	0.96 \pm 0.02	&	1.05 \pm 0.03		&	0.92 \pm 0.03		&	5530 \pm 40		&	4.37 \pm 0.03	&	8.8 \pm 1.9		&	0.26 \pm 0.02	&	(3),(11)	\\
HD 210702	&	1.47 \pm 0.04	&	4.9 \pm 0.1			&	12.9 \pm 0.1		&	4946 \pm 25		&	3.22 \pm 0.02	&	3.1 \pm 0.3		&	-0.05 \pm 0.04	&	(3),(8)	\\
HD 217014	&	1.09 \pm 0.02	&	1.13 \pm 0.03		&	1.34 \pm 0.03		&	5857 \pm 39		&	4.37 \pm 0.02	&	3.8 \pm 1.1		&	0.2 \pm 0.02	&	(3),(11)	\\
HD 217107	&	1.08 \pm 0.01	&	1.11 \pm 0.02		&	1.14 \pm 0.01		&	5676 \pm 31		&	4.38 \pm 0.02	&	4.2 \pm 1		&	0.37 \pm 0.02	&	(3),(11)	\\
HD 217786	&	1.03 \pm 0.02	&	1.27 \pm 0.04		&	1.93 \pm 0.04		&	6031 \pm 55		&	4.23 \pm 0.03	&	6.8 \pm 0.9		&	-0.14 \pm 0.01	&	(3),(15)	\\
HD 218566	&	0.8 \pm 0.01	&	0.77 \pm 0.02		&	0.3 \pm 0.01		&	4880 \pm 16		&	4.57 \pm 0.02	&	8 \pm 3.1		&	0.17 \pm 0.04	&	(3),(15)	\\
HD 219828	&	1.2 \pm 0.04	&	1.58 \pm 0.04		&	2.74 \pm 0.03		&	5921 \pm 53		&	4.11 \pm 0.03	&	5.2 \pm 0.8		&	0.16 \pm 0.04	&	(3),(8)	\\
HD 220074	&	1.2 \pm 0.3		&	49.7 \pm 9.5		&	531.6 \pm 211.7		&	3935 \pm 110	&	1.1 \pm 0.2		&	4.5 \pm 2.8		&	-0.25 \pm 0.25	&	(10)	\\
HD 220773	&	1.154 \pm 0.003	&	1.73 \pm 0.02		&	3.16 \pm 0.01		&	5852 \pm 26		&	4.02 \pm 0.01	&	6.3 \pm 0.1		&	0.11 \pm 0.03	&	(3),(15)	\\
HD 221345	&	1.2 \pm 0.2		&	11 \pm 0.3			&	56 \pm 1			&	4775 \pm 49		&	2.4 \pm 0.1		&	5.6 \pm 3		&	-0.29 \pm 0.03	&	(3),(16)	\\
HD 222155	&	1.05 \pm 0.01	&	1.7 \pm 0.1			&	2.9 \pm 0.1			&	5814 \pm 43		&	4 \pm 0.01		&	8.1 \pm 0.4		&	-0.09 \pm 0.02	&	(3),(16)	\\
HD 231701	&	1.23 \pm 0.01	&	1.48 \pm 0.05		&	2.94 \pm 0.05		&	6211 \pm 71		&	4.18 \pm 0.03	&	3.7 \pm 0.5		&	0.04 \pm 0.02	&	(3),(15)	\\
HD 240210	&	1.241 \pm 0.238	&	19.293 \pm 4.399	&	115.9 \pm 53.5		&	4316 \pm 78		&	1.91 \pm 0.21	&	5.085 \pm 3.089	&	-0.14 \pm 0.03	&	(16)	\\
HD 240237	&	0.614 \pm 0.076	&	0.587 \pm 0.274		&	0.1183 \pm 0.1109	&	4422 \pm 101	&	1.69 \pm 0.24	&	4.42 \pm 4.007	&	-0.24 \pm 0.06	&	(16)
  \enddata
~~(1) {\citet{bensby14}}, (2) {\citet{bonfanti15}}, (3) {\citet{bonfanti16}}, (4) {\citet{fulton16}}, (5) {\citet{gettel12}}, (6) {\citet{giguere15}}, (7) {\citet{gonzalez07}}, (8) {\citet{jofre15}}, (9) {\citet{kang11}}, (10) {\citet{lee13}}, (11) {\citet{maldonado15}}, (12) {\citet{mann15}}, (13) {\citet{massarotti08}}, (14) {\citet{mccarthy14}}, (15) {\citet{santos13}}, (16) {\citet{sousa15}}, (17) {\citet{spina16}}
\end{deluxetable*}

\section{Observations and Data Reduction}
\label{red}

The details on the process undergone to obtain the final reconstructed
images is provided in the previous paper \citep{wittrock16} and with
greater depth in \citet{horch12, horch15} but will be summarized
here. The DSSI obtains the raw speckle data and stores it as FITS data
cubes containing 1000 single short-exposure frames, where each frame
is a $256 \times 256$-pixel image centered on the target. The plate
scales for the observing run at Gemini-North were 0.01081 arcsec/pixel
and 0.01120 arcsec/pixel for the 692~nm and 880~nm channels
respectively. The frames are bias-subtracted, autocorrelated, and then
summed. The result is then Fourier transformed to retrieve the spatial
frequency power spectrum of both the science target and a known
unresolved point source standard. Afterward, the science target's
power spectrum is divided by that of the point source to deconvolve
the effects of the speckle transfer function and obtain a
diffraction-limited estimate of the true power spectrum of the
object. For the raw data frames, the image bispectrum of each frame
has been created (more details on this process has been described in
\citet{lohmann83}). The relaxation algorithm of \citet{meng90} is then
applied to calculate the phase of the object's Fourier transform. This
result is then added with the square root of the deconvolved power
spectrum to arrive at an estimate of the object's Fourier
transform. Next, it is multiplied with a Gaussian low-pass filter of
FWHM width equal to the telescope's diffraction limit. Lastly, we
inverse-Fourier-transform the result to obtain the final reconstructed
image.

\begin{figure*}
  \begin{center}
    \begin{tabular}{cc}
      \includegraphics[angle=90,width=8cm]{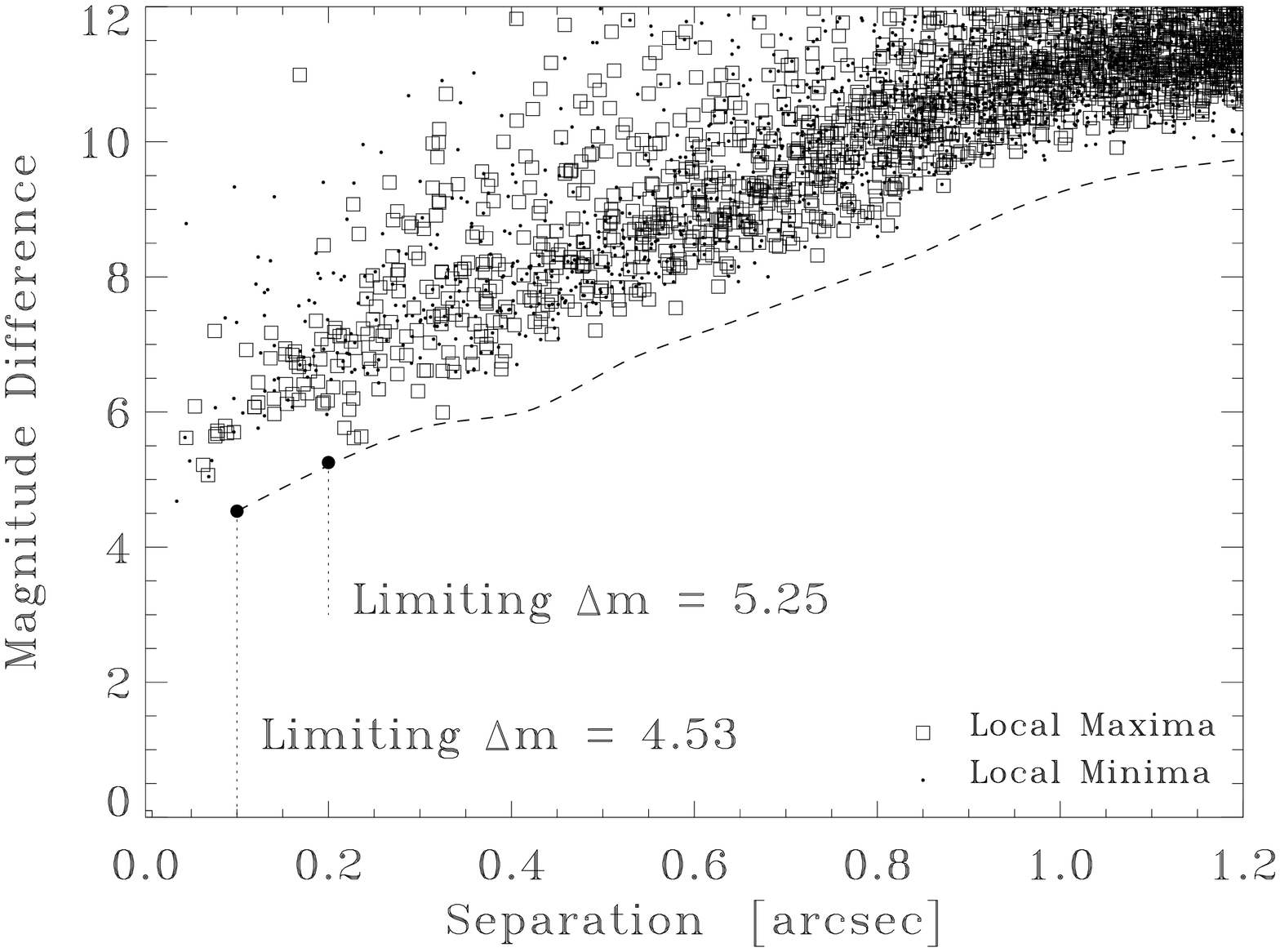} &
      \includegraphics[angle=90,width=8cm]{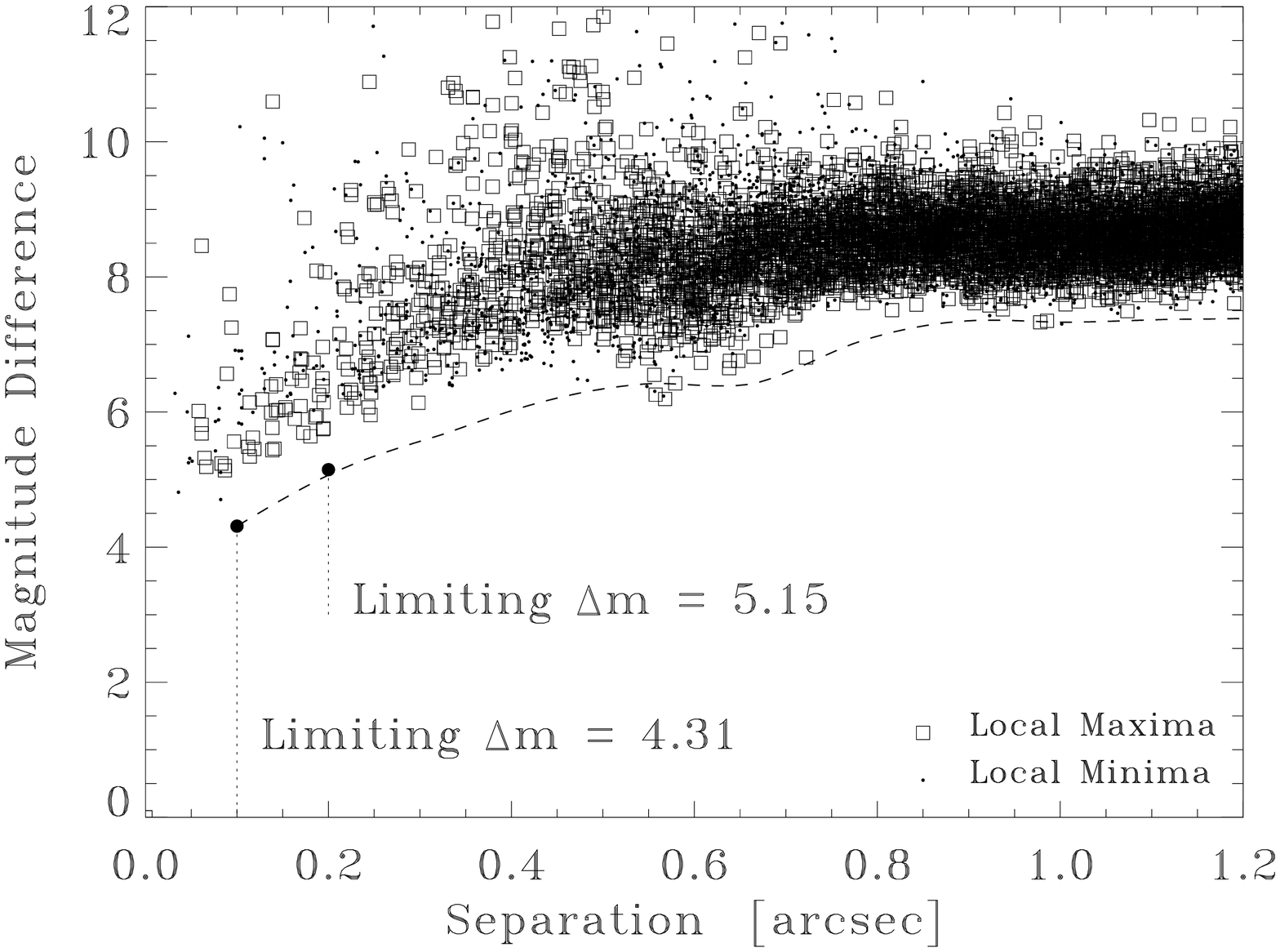} \\
      \includegraphics[angle=90,width=8cm]{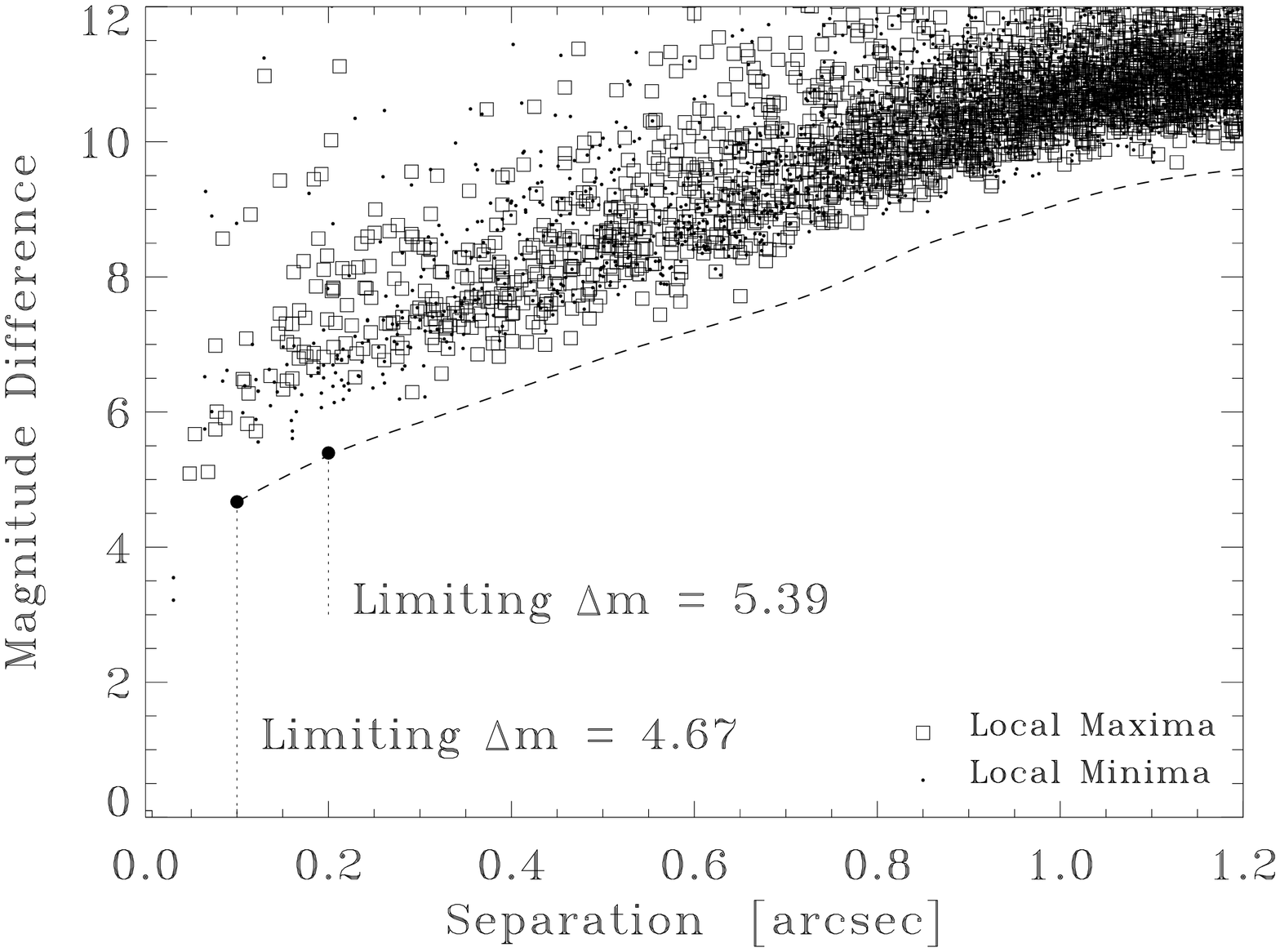} &
      \includegraphics[angle=90,width=8cm]{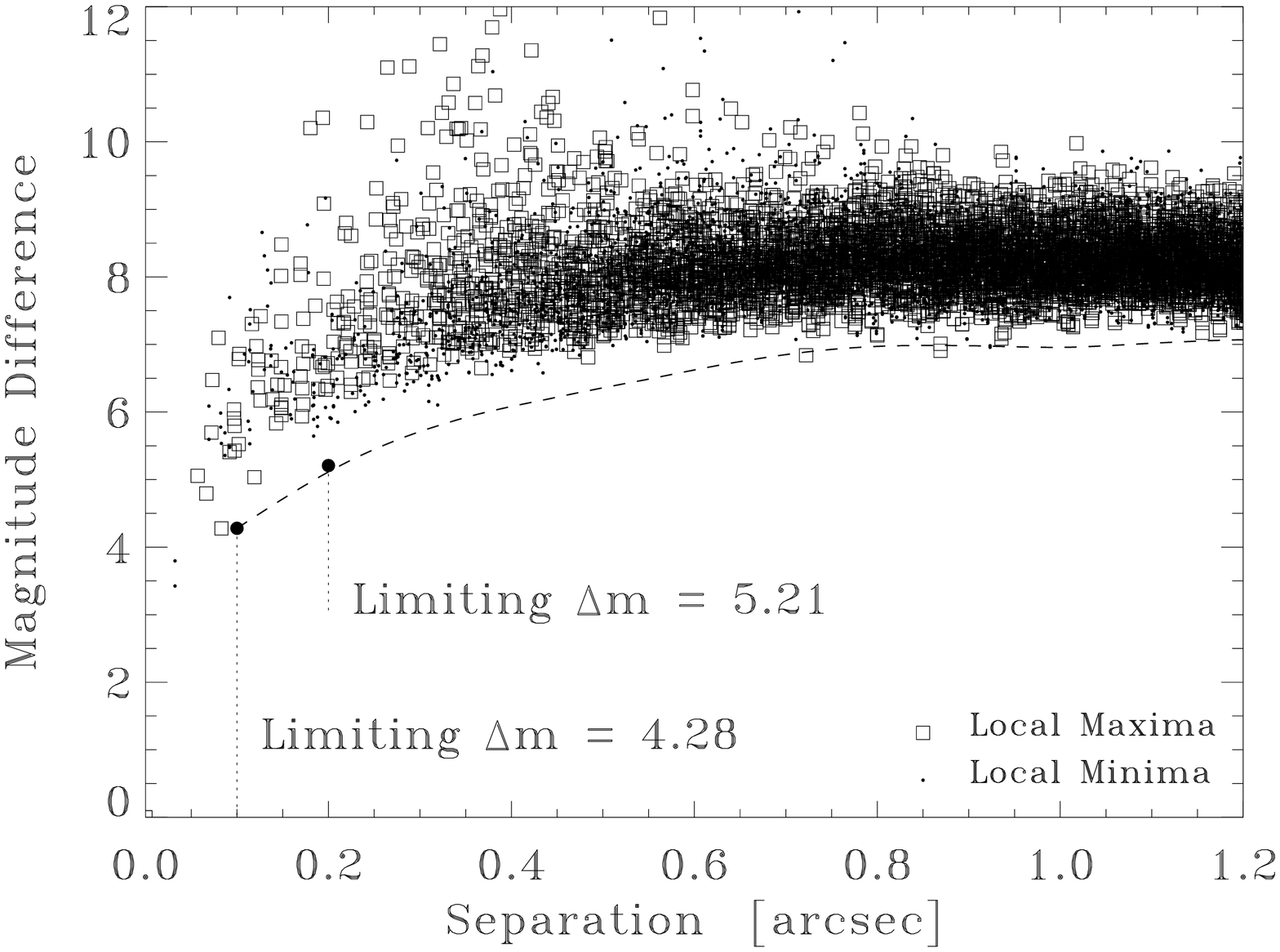} \\
      \includegraphics[angle=90,width=8cm]{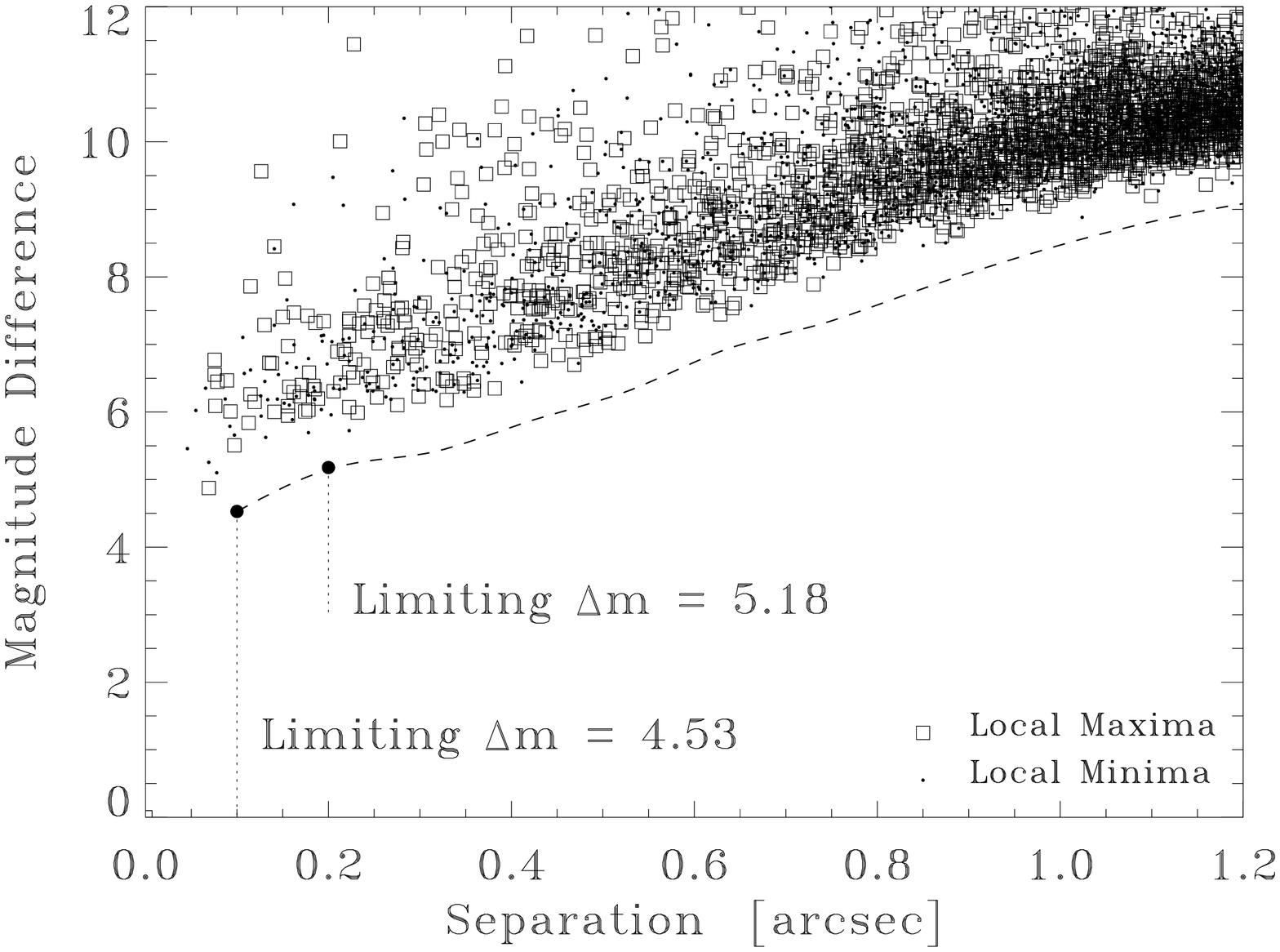} &
      \includegraphics[angle=90,width=8cm]{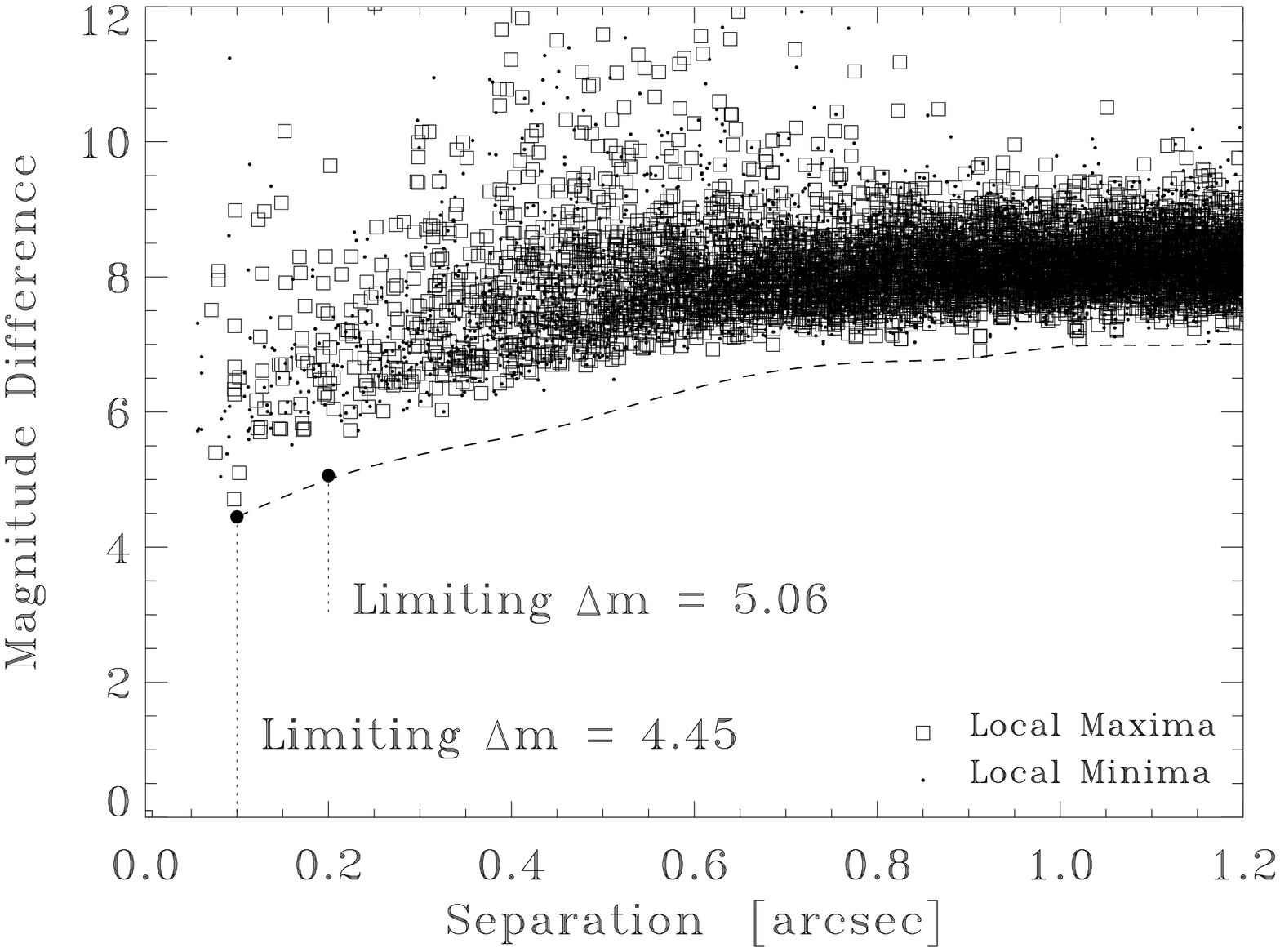}
    \end{tabular}
  \end{center}
  \caption{Limiting magnitude plots for selected targets. Left and
    right columns are sensitivity plots at 692~nm and 880~nm,
    respectively. Each plot shows the limiting magnitude (difference
    between local maxima and minima) as a function of apparent
    separation from a given target in arcsec. The dashed line is a
    cubic spline interpolation of the 5$\sigma$ detection limit. Both
    plots were generated from the corresponding DSSI images. The
    Gemini diffraction limits are 0.021'' and 0.027'' at 692~nm and
    880~nm, respectively.}
  \label{lmplot}
\end{figure*}

With the reconstructed images in hand, we can use the method from
\citet{horch11} to obtain a detection limit curve with respect to
angular separation from the primary star. The average and standard
deviation of the maxima inside annuli are computed to guide our
estimation of the 5-sigma detection limit, which is the mean value
plus five times the standard deviation. These values are in unit of
magnitude difference. The DSSI's diffraction limit on Gemini-North
($0.022''$ at 692 nm and $0.027''$ at 880 nm) constrains the angular
range of annuli from $0.1$ to $\sim$1.2'', and we arbitrarily chose an
increment of $0.1$'' for the annuli. Afterward, we employed a cubic
spline interpolation to achieve a smooth detection limit curve at all
separations in between the two extreme limits. The sensitivity plots
with these curves for some targets are listed in
Figure~\ref{lmplot}. The construction of these sensitivity plots are
described in more detail by \citet{how11}.

\section{Results}
\label{res}

Even though the remaining systems have not yielded the discovery of a
stellar companion like those of HD 2638 and HD 164509
\citep{wittrock16}, such results nevertheless provide an important
contribution to the stellar companion survey. Given that the
field-of-view of the images is $2.8'' \times 2.8''$, a null-detection
implies that the system contains no stellar companions down to the
sensitivity limit of the observation and within the projected size of
the field-of-view. For example, one such system for which we do not
detect a stellar companion and is relatively nearby ($\sim$17~pc) is
14~Her (HD~145675). \citet{wittenmyer07} proposed a second exoplanet,
14~Her~c, with a semi-major axis of 6.9 AU and very low eccentricity
of $0.02 \pm 0.06$. Our non-detection of a stellar companion around
14~Her out to $\sim$25~AU (see Table~\ref{limmag}) indicates that the
observed signal for the outer body is highly likely to be caused by a
planetary body.

Table~\ref{limmag} includes the inner and outer exclusion radii,
limiting magnitudes at $0.1''$ and $0.2''$, the distance moduli, and
the maximum stellar mass of a hypothetical companion for each target,
including HD~2638 and HD~164509 from \citet{wittrock16}. The exclusion
radii are the range of physical separations from the host star that
are observable within the Gemini DSSI's field-of-view. The exclusion
radii and the distance moduli are calculated using the stellar
distances provided in Tables \ref{propi}. The minimum angular
separation is constrained by Gemini's diffraction limits of about
$0.022''$ at 692~nm and $0.027''$ at 880~nm and is $0.05''$. As
mentioned before, the maximum angular separation is $1.2''$, which
provides a constraint on the outer exclusion radius. Thus, the
exclusion radii provide a region where stellar companions with certain
spectral types may be excluded. Therefore, the last two columns tell
us what the maximum mass a stellar companion can have before becoming
detectable via DSSI within the exclusion region. These masses were
calculated using a simple mass-luminosity relationship of $(L_\star /
L_\odot) = (M_\star / M_\odot)^{3.5}$ \citep{kui38} and the given
limiting magnitudes at both 692~nm and 880~nm.

\begin{figure}
  \begin{center}
    \begin{tabular}{c}
      \includegraphics[width=8cm]{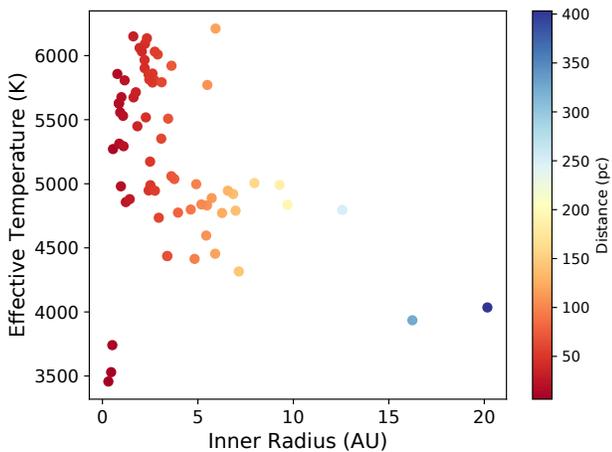}
    \end{tabular}
  \end{center}
  \caption{Inner radius versus effective temperature of the
    targets. The color bar represents the distance in pc.}
  \label{irt}
\end{figure}

\begin{figure*}
  \begin{center}
    \begin{tabular}{cc}
      \includegraphics[width=8cm]{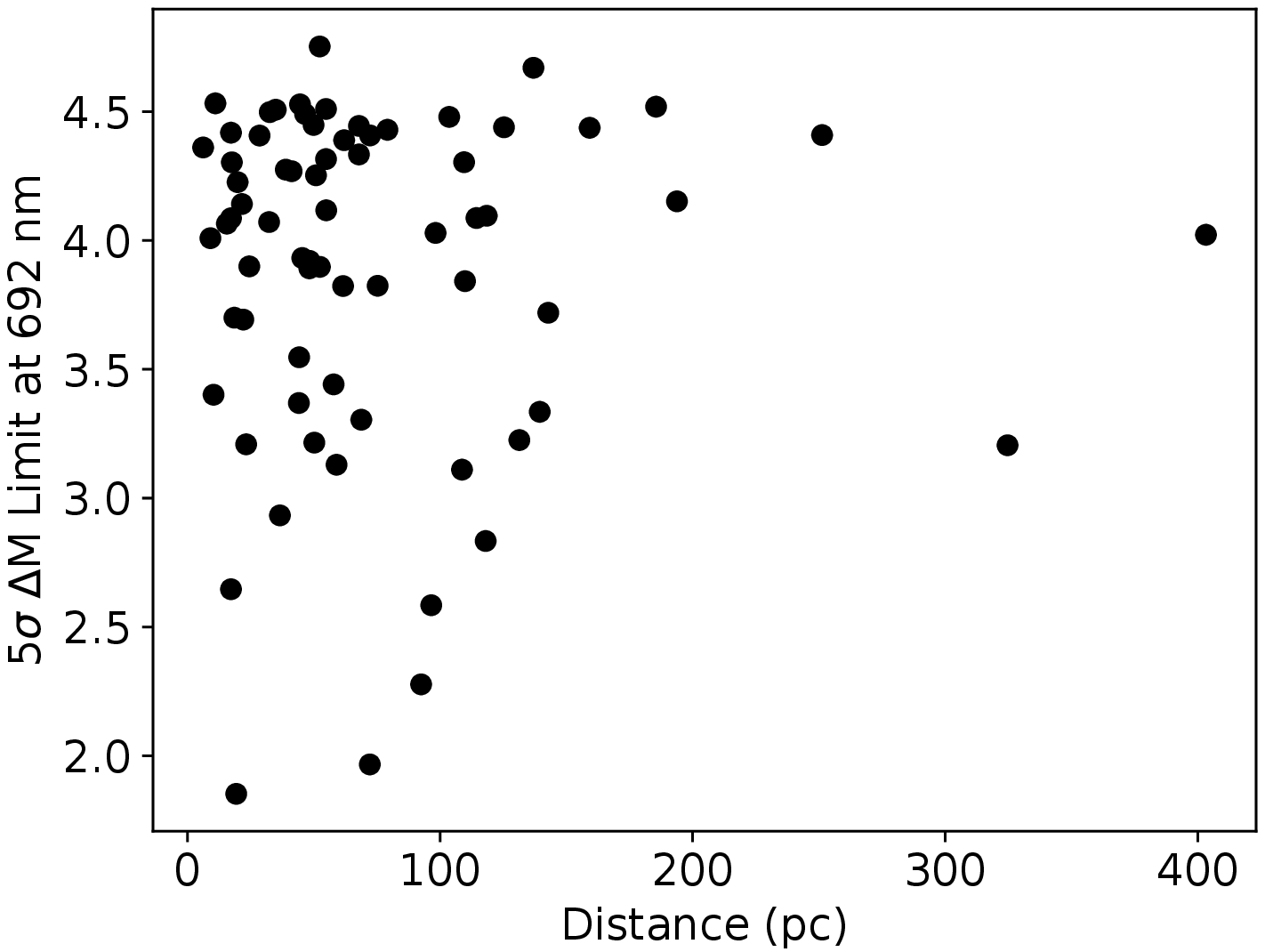} &
      \includegraphics[width=8cm]{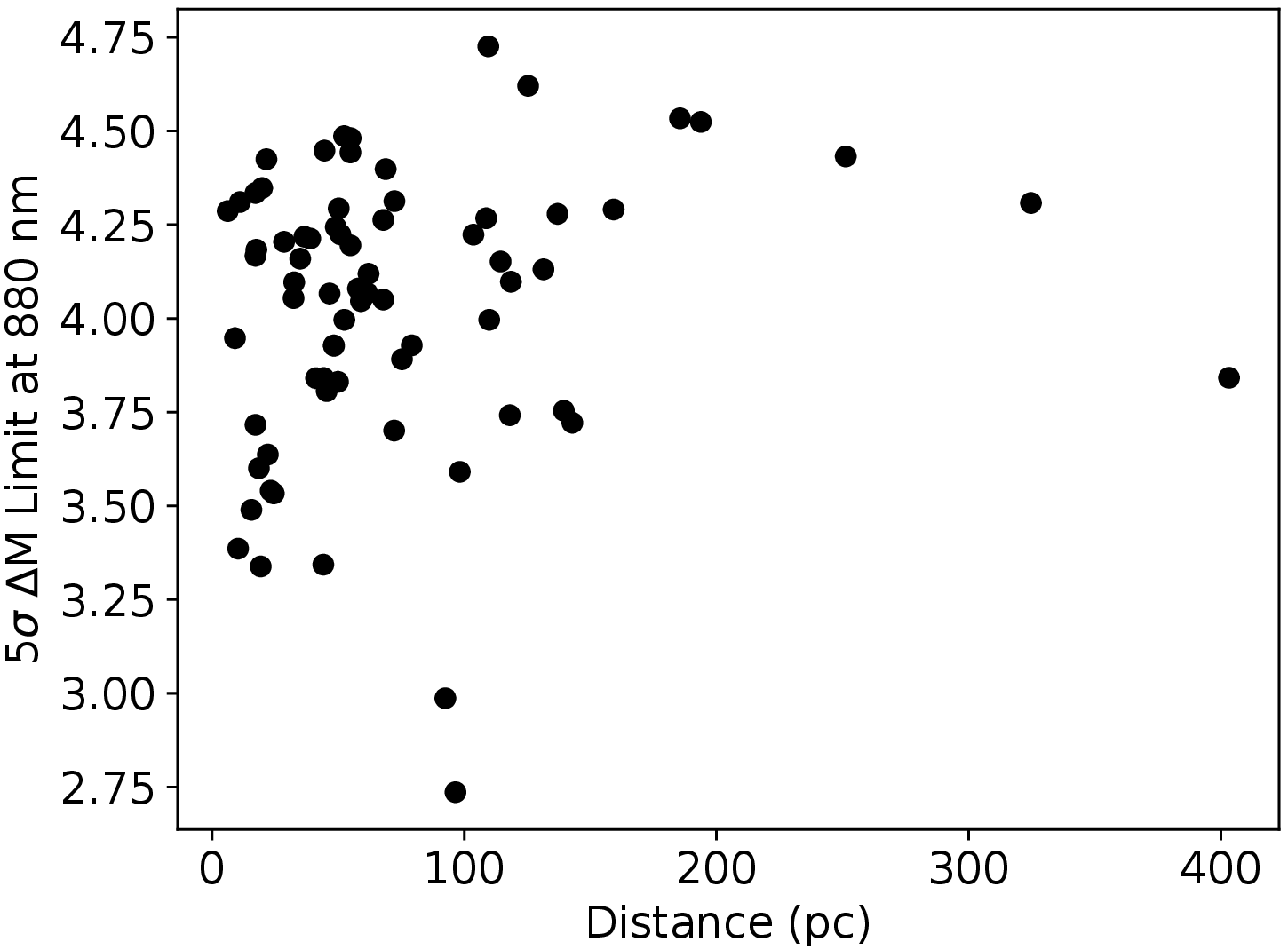}
    \end{tabular}
  \end{center}
  \caption{Correlations between host star distances and limiting
    magnitude at $0.1''$. The plot on the left is at 692~nm while the
    one on the right is at 880~nm.}
  \label{dmag}
\end{figure*}

We constructed three plots, shown in Figure~\ref{irt} and
Figure~\ref{dmag}, using the information from
Table~\ref{limmag}. Figure~\ref{irt} shows some correlation between
the inner exclusion radius and effective temperature of the targets,
with tight clusters at or below 5 AU. The large inner exclusion radius
for some of the cooler stars represents the giant stars in the sample,
as verified by the distance indicators. The plots shown in
Figure~\ref{dmag} show that the limiting magnitude of our observations
are largely consistent between the two passbands used of 692~nm and
880~nm. As for Figure~\ref{tableplot}, BD+48~738, HD~13189, and
HD~240237 were excluded from the plots (see Setion~\ref{obs}).

\begin{deluxetable*}{lCCCCCCCCC}
  \tablecolumns{10}
  \tablewidth{0pt}
  \tablecaption{\label{limmag} Limiting Magnitudes}
  \tablehead{
    Name &
    \multicolumn{2}{c}{Exclusion Radius (AU)} &
    \multicolumn{2}{c}{$5\sigma$ $\Delta$m Limit (692 nm)} &
    \multicolumn{2}{c}{$5\sigma$ $\Delta$m Limit (880 nm)} &
    \colhead{$m-M$} &
    \multicolumn{2}{c}{Max Mass ($M_{\odot}$)} \\
    \colhead{} &
    \colhead{Inner}	&
    \colhead{Outer}	&
    \colhead{0.1''} &
    \colhead{0.2''} &
    \colhead{0.1''} &
    \colhead{0.2''} &
    \colhead{} &
    \colhead{692 nm} &
    \colhead{880 nm}
    }
  \startdata
BD+14 4559	&	2.42	&	67.70	&	3.92	&	4.32	&	3.93	&	4.62	&	3.42	&	0.26	&	0.26	\\
BD+48 738	&	17.54	&	491.23	&	3.09	&	5.01	&	3.72	&	4.74	&	7.73	&	1.35	&	1.14	\\
GJ 581		&	0.31	&	8.70	&	4.36	&	5.14	&	4.29	&	5.23	&	-1.03	&	0.09	&	0.09	\\
GJ 649		&	0.52	&	14.48	&	3.40	&	4.23	&	3.39	&	4.27	&	0.07	&	0.17	&	0.17	\\
GJ 849		&	0.45	&	12.73	&	4.01	&	4.55	&	3.95	&	4.55	&	-0.21	&	0.13	&	0.13	\\
HD 1461		&	1.16	&	32.54	&	3.21	&	4.15	&	3.54	&	4.89	&	1.83	&	0.45	&	0.41	\\
HD 1502		&	7.96	&	222.93	&	4.44	&	5.20	&	4.29	&	4.97	&	6.01	&	0.63	&	0.65	\\
HD 2638		&	2.50	&	69.90	&	4.45	&	4.96	&	3.83	&	4.63	&	3.49	&	0.24	&	0.28	\\
HD 3651		&	0.55	&	15.48	&	4.53	&	5.25	&	4.31	&	5.15	&	0.22	&	0.25	&	0.27	\\
HD 4313		&	6.85	&	191.78	&	4.67	&	5.39	&	4.28	&	5.21	&	5.68	&	0.62	&	0.69	\\
HD 5319		&	5.72	&	160.18	&	4.09	&	5.30	&	4.15	&	5.23	&	5.29	&	0.62	&	0.61	\\
HD 5891		&	12.56	&	351.76	&	4.41	&	5.37	&	4.43	&	5.31	&	7.00	&	0.89	&	0.89	\\
HD 6718		&	2.74	&	76.80	&	4.32	&	4.80	&	4.20	&	4.95	&	3.70	&	0.33	&	0.34	\\
HD 7449		&	1.95	&	54.50	&	4.28	&	4.91	&	4.21	&	4.80	&	2.95	&	0.35	&	0.35	\\
HD 8574		&	2.23	&	62.39	&	4.53	&	5.18	&	4.45	&	5.06	&	3.24	&	0.39	&	0.40	\\
HD 9446		&	2.62	&	73.30	&	4.75	&	5.24	&	4.49	&	5.08	&	3.59	&	0.29	&	0.31	\\
HD 10697	&	1.63	&	45.60	&	4.50	&	5.20	&	4.10	&	4.80	&	2.56	&	0.41	&	0.46	\\
HD 12661	&	1.75	&	48.93	&	4.51	&	5.22	&	4.16	&	5.15	&	2.72	&	0.32	&	0.35	\\
HD 13189	&	28.09	&	786.52	&	4.03	&	5.33	&	3.73	&	4.87	&	8.75	&	0.00	&	0.00	\\
HD 13931	&	2.21	&	61.92	&	3.55	&	5.05	&	3.84	&	4.97	&	3.23	&	0.44	&	0.41	\\
HD 16175	&	2.89	&	81.02	&	3.44	&	5.03	&	4.08	&	5.30	&	3.81	&	0.57	&	0.48	\\
HD 16400	&	4.63	&	129.51	&	2.28	&	4.22	&	2.99	&	3.66	&	4.83	&	1.77	&	1.47	\\
HD 16760	&	2.27	&	63.64	&	3.93	&	4.99	&	3.81	&	4.77	&	3.29	&	0.30	&	0.31	\\
HD 17092	&	5.43	&	152.17	&	3.11	&	4.64	&	4.27	&	5.07	&	5.18	&	1.30	&	0.96	\\
HD 136118	&	2.33	&	65.21	&	4.49	&	5.37	&	4.07	&	5.07	&	3.34	&	0.42	&	0.47	\\
HD 136418	&	4.91	&	137.52	&	4.03	&	4.99	&	3.59	&	4.52	&	4.96	&	0.60	&	0.67	\\
HD 137510	&	2.06	&	57.76	&	4.27	&	4.92	&	3.84	&	4.83	&	3.08	&	0.49	&	0.55	\\
HD 139357	&	5.90	&	165.29	&	2.83	&	4.73	&	3.74	&	4.65	&	5.36	&	1.62	&	1.28	\\
HD 142245	&	5.48	&	153.34	&	4.30	&	5.01	&	4.73	&	5.30	&	5.20	&	0.67	&	0.60	\\
HD 143107	&	3.39	&	95.04	&	4.33	&	5.32	&	4.05	&	5.25	&	4.16	&	1.34	&	1.44	\\
HD 143107	&	3.39	&	95.04	&	4.44	&	5.19	&	4.26	&	5.20	&	4.16	&	1.30	&	1.37	\\
HD 143761	&	0.86	&	24.13	&	2.65	&	3.66	&	4.17	&	4.96	&	1.18	&	0.58	&	0.39	\\
HD 143761	&	0.86	&	24.13	&	4.09	&	4.81	&	3.72	&	4.68	&	1.18	&	0.40	&	0.44	\\
HD 143761	&	0.86	&	24.13	&	4.42	&	4.95	&	4.33	&	5.19	&	1.18	&	0.36	&	0.37	\\
HD 145457	&	6.27	&	175.44	&	4.44	&	4.96	&	4.62	&	5.15	&	5.49	&	0.90	&	0.86	\\
HD 145675	&	0.88	&	24.60	&	4.30	&	5.13	&	4.18	&	5.34	&	1.22	&	0.28	&	0.29	\\
HD 149143	&	3.10	&	86.85	&	4.39	&	4.96	&	4.12	&	5.09	&	3.96	&	0.39	&	0.42	\\
HD 152581	&	9.28	&	259.74	&	4.52	&	5.09	&	4.53	&	5.05	&	6.34	&	0.67	&	0.67	\\
HD 154345	&	0.93	&	26.02	&	3.70	&	4.96	&	3.60	&	4.72	&	1.35	&	0.33	&	0.34	\\
HD 155358	&	2.21	&	61.76	&	3.37	&	4.75	&	3.34	&	4.55	&	3.22	&	0.51	&	0.51	\\
HD 156279	&	1.83	&	51.24	&	2.93	&	4.53	&	4.22	&	4.86	&	2.82	&	0.42	&	0.30	\\
HD 156668	&	1.22	&	34.26	&	3.90	&	5.18	&	3.53	&	4.71	&	1.94	&	0.25	&	0.27	\\
HD 158038	&	5.18	&	145.08	&	4.48	&	5.24	&	4.22	&	5.07	&	5.08	&	0.62	&	0.67	\\
HD 163607	&	3.44	&	96.35	&	3.30	&	4.51	&	4.40	&	4.95	&	4.19	&	0.55	&	0.41	\\
HD 164509	&	2.62	&	73.41	&	3.90	&	4.15	&	4.00	&	4.52	&	3.60	&	0.39	&	0.38	\\
HD 164922	&	1.11	&	30.97	&	3.69	&	4.88	&	3.64	&	4.76	&	1.72	&	0.34	&	0.35	\\
HD 167042	&	2.51	&	70.32	&	3.22	&	3.84	&	4.29	&	5.15	&	3.50	&	0.84	&	0.64	\\
HD 170693	&	4.83	&	135.14	&	2.58	&	4.05	&	2.74	&	3.84	&	4.92	&	2.10	&	2.02	\\
HD 171028	&	5.49	&	153.85	&	3.84	&	4.09	&	4.00	&	4.45	&	5.20	&	0.54	&	0.52	\\
HD 173416	&	6.97	&	195.26	&	3.33	&	3.51	&	3.75	&	4.01	&	5.72	&	1.45	&	1.30	\\
HD 177830	&	2.95	&	82.64	&	3.13	&	3.23	&	4.05	&	4.29	&	3.86	&	0.71	&	0.56	\\
HD 180314	&	6.57	&	183.97	&	3.23	&	3.48	&	4.13	&	4.46	&	5.59	&	1.23	&	0.97	\\
HD 187123	&	2.41	&	67.57	&	3.89	&	3.98	&	3.93	&	4.34	&	3.42	&	0.40	&	0.39	\\
HD 190228	&	3.08	&	86.26	&	3.82	&	4.03	&	4.07	&	4.56	&	3.95	&	0.56	&	0.52	\\
HD 192263	&	0.97	&	27.04	&	1.85	&	3.55	&	3.34	&	4.21	&	1.43	&	0.44	&	0.29	\\
HD 197037	&	1.62	&	45.26	&	4.07	&	4.25	&	4.05	&	4.63	&	2.55	&	0.39	&	0.39	\\
HD 199665	&	3.77	&	105.42	&	3.82	&	4.30	&	3.89	&	4.48	&	4.38	&	1.01	&	0.99	\\
HD 200964	&	3.61	&	101.08	&	1.97	&	3.18	&	3.70	&	4.60	&	4.29	&	1.23	&	0.78	\\
HD 206610	&	9.69	&	271.32	&	4.15	&	4.60	&	4.52	&	5.00	&	6.44	&	0.77	&	0.69	\\
HD 208527	&	20.16	&	564.52	&	4.02	&	4.17	&	3.84	&	4.10	&	8.03	&	2.18	&	2.29	\\
HD 210277	&	1.08	&	30.19	&	4.14	&	4.27	&	4.42	&	5.05	&	1.67	&	0.33	&	0.30	\\
HD 210702	&	2.75	&	76.92	&	4.12	&	4.54	&	4.48	&	4.74	&	3.70	&	0.70	&	0.64	\\
HD 217014	&	0.78	&	21.85	&	4.07	&	4.60	&	3.49	&	4.24	&	0.97	&	0.37	&	0.43	\\
HD 217107	&	0.99	&	27.80	&	4.23	&	4.88	&	4.35	&	5.27	&	1.49	&	0.34	&	0.33	\\
HD 217786	&	2.74	&	76.80	&	4.51	&	5.20	&	4.44	&	5.16	&	3.70	&	0.37	&	0.37	\\
HD 218566	&	1.43	&	39.98	&	4.41	&	5.19	&	4.20	&	5.18	&	2.28	&	0.22	&	0.23	\\
HD 219828	&	3.62	&	101.23	&	4.41	&	5.19	&	4.31	&	5.12	&	4.30	&	0.42	&	0.43	\\
HD 220074	&	16.23	&	454.55	&	3.21	&	3.89	&	4.31	&	4.67	&	7.56	&	2.58	&	1.93	\\
HD 220773	&	2.54	&	71.25	&	4.25	&	4.94	&	4.22	&	4.87	&	3.53	&	0.45	&	0.46	\\
HD 221345	&	3.96	&	110.85	&	4.43	&	5.00	&	3.93	&	4.87	&	4.49	&	0.98	&	1.12	\\
HD 222155	&	2.45	&	68.69	&	3.91	&	4.15	&	4.24	&	4.61	&	3.45	&	0.48	&	0.44	\\
HD 231701	&	5.92	&	165.88	&	4.10	&	4.18	&	4.10	&	4.31	&	5.37	&	0.46	&	0.46	\\
HD 240210	&	7.14	&	200.00	&	3.72	&	4.13	&	3.72	&	4.25	&	5.77	&	1.46	&	1.46	\\
HD 240237	&	263.16	&	7368.42	&	3.30	&	4.65	&	4.52	&	5.18	&	13.61	&	0.23	&	0.17
  \enddata
\end{deluxetable*}

\section{Conclusion}
\label{con}

Detection of stellar companions to exoplanet host stars is now
relatively common. Their discovery, or lack of it, are of beneficial
contributions to the structure of the stellar systems since the
presence of a stellar companion significantly influences the orbital
dynamics and models for formation processes. For planets discovered
using the RV technique, the search for stellar companions plays a major
role in the correct interpretation of residual RV trends present in
the data. Furthermore, determining if any of the known exoplanet host
stars are single, binary, or multiple systems is absolutely essential
in avoiding situations where a gas giant would be mistaken for a
terrestrial or even an Earth-like planet.

Our DSSI survey monitored 71 stars for which 2 were detected to have
stellar companions \citep{wittrock16} and the remaining 69 show no
evidence of stellar companions within the $2.8'' \times 2.8''$
field-of view and above the instrument's sensitivity limit. This
increases the probability that remaining detected objects in the RV
data, if any, are planetary bodies if not extremely low-mass stars or
brown dwarfs. An example of this is the 14~Her system, for which the
partial phase coverage of the RV signal detected by
\citet{wittenmyer07} is better explained by a planetary body rather
than stellar, since our exclusion range out to $\sim$25~AU completely
encompasses the postulated semi-major axis of the proposed
14~Her~c. The exclusion radii listed in Table~\ref{limmag} provide a
physical range of separations from each star within which any future
RV detection of objects gravitationally bound to these stars may now
be more closely associated with a planetary object.


\section{Acknowledgments}

The authors would like to thank the referee for providing feedback
that improved the quality of the paper. This work is based on
observations obtained at the Gemini Observatory, which is operated by
the Association of Universities for Research in Astronomy, Inc., under
a cooperative agreement with the NSF on behalf of the Gemini
partnership: the National Science Foundation (United States), the
National Research Council (Canada), CONICYT (Chile), the Australian
Research Council (Australia), Minist\'{e}rio da Ci\^{e}ncia,
Tecnologia e Inova\c{c}\~{a}o (Brazil) and Ministerio de Ciencia,
Tecnolog\'{i}a e Innovaci\'{o}n Productiva (Argentina). This research
has made use of the NASA Exoplanet Archive, which is operated by the
California Institute of Technology, under contract with the National
Aeronautics and Space Administration under the Exoplanet Exploration
Program. The results reported herein benefited from collaborations
and/or information exchange within NASA's Nexus for Exoplanet System
Science (NExSS) research coordination network sponsored by NASA's
Science Mission Directorate.


\end{document}